\documentclass[11pt]{article}

\usepackage{charter}
\usepackage[charter,cal=cmcal]{mathdesign}

\usepackage[margin=1in]{geometry}
\usepackage{microtype}
\usepackage{textcomp}

\usepackage{color}
\definecolor{darkblue}{rgb}{0,0,0.5}
\definecolor{darkgreen}{rgb}{0,0.5,0}
\usepackage[colorlinks=true,linkcolor=darkblue,urlcolor=darkblue,citecolor=darkgreen]{hyperref}

\usepackage{graphicx}

\newtheorem{theorem}{Theorem}

\usepackage[absolute]{textpos}

\title{\bf The Case for Quantum Key Distribution}

\author{Douglas Stebila$^{1}$, Michele Mosca$^{2,3,4}$, and Norbert L\"utkenhaus$^{2,4,5}$ \\\small
\small \ \\
{\small $^{1}$ \quad \it Information Security Institute, Queensland University of Technology, Brisbane, Australia}
\\ {\small $^{2}$ \quad \it Institute for Quantum Computing, University of Waterloo}
\\ {\small $^{3}$ \quad \it Dept. of Combinatorics \& Optimization, University of Waterloo}
\\ {\small $^{4}$ \quad \it Perimeter Institute for Theoretical Physics}
\\ {\small $^{5}$ \quad \it Dept. of Physics \& Astronomy, University of Waterloo}
\\ {\small \it Waterloo, Ontario, Canada} \\
\small Email: {\tt \href{mailto:douglas@stebila.ca}{douglas@stebila.ca}, \href{mailto:mmosca@iqc.ca}{mmosca@iqc.ca}, \href{mailto:nlutkenhaus@iqc.ca}{nlutkenhaus@iqc.ca}}
}

\date{December 2, 2009}

\begin{document}

\maketitle

\begin{abstract}
Quantum key distribution (QKD) promises secure key agreement by using quantum mechanical systems.  We argue that QKD will be an important part of future cryptographic infrastructures.  It can provide long-term confidentiality for encrypted information without reliance on computational assumptions.  Although QKD still requires authentication to prevent man-in-the-middle attacks, it can make use of either information-theoretically secure symmetric key authentication or computationally secure public key authentication: even when using public key authentication, we argue that QKD still offers stronger security than classical key agreement.
\end{abstract}

\section{Introduction}\label{sec:intro}

Since its discovery, the field of quantum cryptography --- and in particular, quantum key distribution (QKD) --- has garnered widespread technical and popular interest.  The promise of ``unconditional security'' has brought public interest, but the often unbridled optimism expressed for this field has also spawned criticism and analysis \cite{Sch03,PPS04,Sch07,Sch08}.

QKD is a new tool in the cryptographer's toolbox: it allows for secure key agreement over an untrusted channel where the output key is entirely independent from any input value, a task that is impossible using classical\footnote{All computation must be viewed as taking place in a physical system described by particular laws of nature.  By \emph{classical cryptography}, we mean cryptography taking place in a computational and communication system modelled with classical physics (i.e., non-quantum-mechanical and non-relativistic physics); that is, using processes described by probabilistic Turing machines.} cryptography.  QKD does not eliminate the need for other cryptographic primitives, such as authentication, but it can be used to build systems with new security properties.  As experimental research continues, we expect the costs and challenges of using QKD to decrease to the point where QKD systems can be deployed affordably and their behaviour can be certified.

Through the rest of this paper, we restrict our discussion on quantum cryptography to quantum key distribution (QKD).  Many other quantum cryptographic primitives exist --- quantum private channels, quantum public key encryption, quantum coin tossing, blind quantum computation, quantum money --- but almost all require a medium- to large- scale quantum computer for implementation.  QKD, on the other hand, has already been implemented by many different groups, has seen attempts at commercialization, and thus its potential role in upcoming security infrastructures merits serious examination.

There are three phases (which are sometimes intertwined) to establishing secure communications:
\begin{enumerate}
\item \emph{Key agreement:} Two parties agree upon a secure, shared private key.  
\item \emph{Authentication:} Allows a party to be certain that a message comes from a particular party.  In order for key agreement to avoid man-in-the-middle attacks, authentication of some form must be used.
\item \emph{Key usage:} Once a secure key is established, it can be used for encryption (using a one-time pad or some other cipher), further authentication, or other cryptographic purposes.
\end{enumerate}
QKD is just one part of this overall information security infrastructure: two parties can agree upon a private key, the security of which depends on no computational assumptions, and which is entirely independent of any input to the protocol.

If we live in a world where we can reasonably expect public key cryptography to be secure in the short- to medium-term, then the combination of public key cryptography for authentication and QKD for key agreement can lead to very strong long-term security with all the convenience and benefits we have come to expect from distributed authentication in a public key infrastructure.

If we live in a world where public key cryptography can no longer be employed safely, we must revert to doing classical key establishment over a private channel, such as a trusted courier, or use QKD.  QKD would still require a private channel to establish authentication keys.  Instead of just establishing short authentication keys, a private channel could in principle be used to exchange an amount of key comparable to what QKD could produce over a long period of time.  However, in this setting QKD can have an advantage because the amount of private communication required is much less and because the session keys output by QKD are independent from the keys transmitted across the private channel, leaving a short time window in which compromised keying material can affect the security of future sessions.  How much of an advantage this is in practice will depend on the nature of the private channel in question and the trust assumptions.

If we live in a world where there exist public key agreement schemes that are believed to be secure indefinitely, then there is a reduced case for QKD, but it is still of interest for a variety of reasons.  QKD creates random, independent session keys, which can reduce the damage caused by ephemeral key leakage.  Other forms of quantum cryptography may also be of interest, especially for the secure communication of quantum information if quantum computing becomes widespread.

Experimental research on quantum key distribution continues to improve the usability, rate, and distance of QKD systems, and the ability to provide and certify their physical security.  As public key cryptography systems are retooled with new algorithms and standards over the coming years, there is an opportunity to incorporate QKD as a new tool offering fundamentally new security features.

\paragraph{Related work.}  This work is motivated as a response to other opinions about the role of QKD, especially the thoughtful note ``Why quantum cryptography?'' by Paterson, Piper, and Schack \cite{PPS04}.  Our discussion on encryption and authentication addresses many of the same points as \cite{PPS04} with an optimistic view of the prospect of post-quantum public key cryptography; we provide additional information on the assumptions for the security of QKD, the current state of QKD implementations, and how the structure of QKD networks will evolve as technology progresses.  A response by the SECOQC project \cite{SECOQC07} addresses related concerns as well, with special attention paid to the networks of QKD links.

\paragraph{Outline.}  In the rest of this paper, we argue that QKD has a valuable role to play in future security infrastructures.  In Section~\ref{sec:qkd}, we give an overview of how QKD works, and give an example where its high security is needed in Section~ \ref{sec:usecase}.  We describe the conditions for the security of QKD in Section~ \ref{sec:security}.  We then discuss the other parts of the communication infrastructure: encryption in Section~ \ref{sec:enc} and authentication in Section~ \ref{sec:auth}. In Section~\ref{sec:limits}, we discuss some limitations to QKD as it stands and how they may be overcome, with special consideration to networks of QKD devices in Section~\ref{sec:networks}.  We offer a concluding statement in Section~\ref{sec:concl}.

\section{A Brief Introduction to QKD}\label{sec:qkd}

In this section we provide a very brief overview of quantum key distribution.  More detailed explanations are available from a variety of sources \cite{NC00,SECOQC07,SBCDLP08}.

In QKD, two parties, Alice and Bob, obtain some quantum states and measure them.  They communicate (all communication from this point onwards is classical) to determine which of their measurement results could lead to secret key bits; some are discarded in a process called sifting because the measurement settings were incompatible.  They perform error correction and then estimate a security parameter which describes how much information an eavesdropper might have about their key data.  If this amount is above a certain threshold, then they abort as they cannot guarantee any secrecy whatsoever.  If it is below the threshold, then they can apply privacy amplification to squeeze out any remaining information the eavesdropper might have, and arrive at a shared secret key.  Some of this classical communication must be authenticated to avoid man-in-the-middle attacks.  Some portions of the protocol can fail with negligible probability.

\begin{figure}
\begin{center}
\includegraphics[width=4.5in]{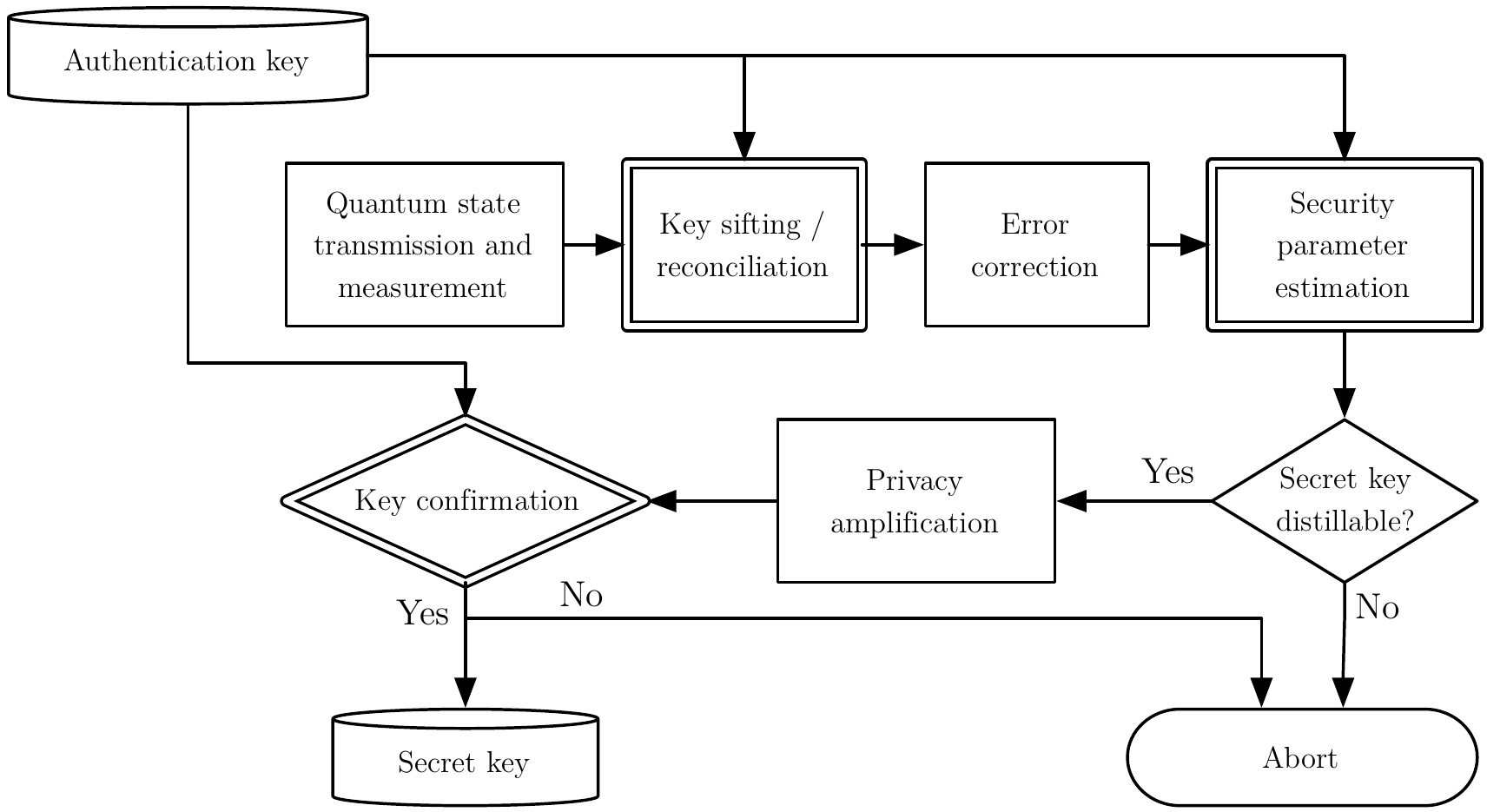}
\end{center}
\caption{Flow chart of the stages of a quantum key distribution protocol.  Stages with double lines require classical authentication.}\label{fig:qkd:flowchart}
\end{figure}

A flow chart describing the stages of quantum key distribution is given in Figure~\ref{fig:qkd:flowchart}.

Once a secret key has been established by QKD, it can be used for a variety of purposes.  The most common approach is to use it as the secret key in a one-time pad to achieve unconditionally secure encryption.  The key can also be used for classical authentication in subsequent rounds of QKD.

We can expect that as QKD research continues, QKD devices will become more robust, easier to configure, less expensive, and smaller, perhaps sufficiently miniaturized to fit on a single circuit board.

\section{Who Needs Quantum Key Distribution?}\label{sec:usecase}

It is widely understood that ``security is a chain; it's as strong as the weakest link'' \cite{Sch03}, and cryptography, even public key cryptography, is indeed  one of the strongest links in the chain.  We cannot trust that a particular computationally secure cryptographic scheme and parameter size will remain secure indefinitely, and many expert recommendations are unwilling to provide guidance for much more than 30 years in the future.  While much of the information being encrypted today does not need 30 years of security, some does.  

Moreover, it is important to plan well in advance for changes in security technology.  Suppose, for example, that a particular application using RSA or elliptic curve cryptography (ECC) needs information to be secure for $x$ years, and it takes $y$ years to retool the infrastructure to a new cryptosystem.  If large-scale quantum computers capable of breaking RSA or ECC are built within $z$ years, with $z < x+y$, then we are already too late: we need to start planning to use new cryptosystems long before old ones are broken.

Government, military, and intelligence agencies need long-term security.  For example, the UK government did not declassify the 1945 report on its efforts in breaking the Tunny cipher during World War II until 2000 \cite{GMT45}, and the US government's current classification regime keeps documents classified for up to 25 years \cite[\S 1.5(b)]{Bus03}.

Businesses trying to protect long-term strategic trade secrets may also wish for long-term confidentiality.  Situations with long-term deployments but well-specified communication requirements could also benefit from QKD: it is inconvenient and expensive to have to upgrade the 1.5 million automated teller machines (ATMs) worldwide whenever the latest cryptographic protocol is broken or deemed obsolete, but QKD could provide standards less likely to change due to cryptanalysis.

One particular industry likely to require long-term, future-proof security is health care.  Health care systems are slowly but irreversibly becoming more electronic, and health care records need privacy for 100 years or more.  Securing the storage of these records in data centers is essential, of course, and quantum key distribution does not aim to solve this difficult problem.  Equally important, however, is the secure communication of health care records, which can be protected by the information-theoretic security offered by quantum key distribution.

Quantum key distribution is also not the only way to establish information theoretically secure keys.  The physical transfer of long, randomly generated keys is also an information theoretically secure key distribution scheme.  With hard drive prices approaching US \$0.10 per gigabyte, one should not underestimate ``the bandwidth of a truck filled with hard drives'' (although increases in fuel prices may counteract the cost efficiency of such a communication system).  This approach is not appropriate for all scenarios.  In some cases, it may be impossible to rekey a system in this manner (e.g., satellites and space probes).  It requires assurances that the physical keys were transported securely.  It also requires secure storage of large amounts of key until use.  QKD requires only a small amount of key, the authentication key, to be securely stored until use.  Importantly, QKD can generate fresh encryption keys on demand that need only be stored for the short time period between key generation and message encryption/decryption, rather than needing large secure key storage since the distribution of the systems.

Moreover, research into experimental quantum information is still at such an early stage that one cannot predict the final form of the products that could be developed from this technology, and these systems may come to exceed the expectations and dreams of today's researchers and engineers.

\section{The Security of QKD}\label{sec:security}

Quantum key distribution is often described by its proponents as ``unconditionally secure'' to emphasize its difference with computationally secure classical cryptographic protocols.  While there are still conditions that need to be satisfied for quantum key distribution to be secure, the phrase ``unconditionally secure'' is justified because, not only are the conditions reduced, they are in some sense minimal necessary conditions.  Any secure key agreement protocol must make a few minimal assumptions, for security cannot come from nothing: we must be able to identify and authenticate the communicating parties, we must be able to have some private location to perform local operations, and all parties must operate within the laws of physics.

The following statement describes the security of quantum key distribution, and there are many formal mathematical arguments for the security of QKD (e.g., \cite{May97,LC99,GLLP04}).

\begin{theorem}[Security statement for quantum key distribution]
If 
\vspace{-1.5mm}
\begin{enumerate}
\setlength{\itemsep}{-1mm}
\item[A1)] quantum mechanics is correct, and
\item[A2)] authentication is secure, and
\item[A3)] our devices are reasonably secure,
\end{enumerate}
\vspace{-1.5mm}
then with high probability the key established by quantum key distribution is a random secret key independent (up to a negligible difference) of input values.
\end{theorem}

\paragraph{Assumption 1: Quantum mechanics is correct.}  
This assumption requires that any eavesdropper be bounded by the laws of quantum mechanics, although within this realm there are no further restrictions beyond the eavesdropper's inability to access the devices.  In particular, we allow the eavesdropper to have arbitrarily large quantum computing technology, far more powerful than the current state of the art.  Quantum mechanics has been tested experimentally for nearly a century, to very high precision.  But even if quantum mechanics is superseded by a new physical theory, it is not necessarily true that quantum key distribution would be insecure: for example, secure key distribution can be achieved in a manner similar to QKD solely based on the assumption that no faster-than-light communication is possible \cite{BHK05}.
 
\paragraph{Assumption 2: Authentication is secure.}
This assumption is one of the main concerns of those evaluating quantum key distribution.  In order to be protected against man-in-the-middle attacks, much of the classical communication in QKD must be authenticated.  Authentication can be achieved with unconditional security using short shared keys, or with computational security using public key cryptography.  We discuss the issue of authentication in greater detail in Section~\ref{sec:auth}.

\paragraph{Assumption 3: Our devices are secure.}
Constructing a QKD implementation that is verifiably secure is a substantial engineering challenge that researchers are still working on.  Although the first prototype QKD system leaked key information over a side channel (it made different noises depending on the photon polarization, and thus the ``prototype was unconditionally secure against any eavesdropper who happened to be deaf'' \cite{Bra06}), experimental cryptanalysis leads to better theoretical and practical security.  More sophisticated side-channel attacks continue to be proposed against particular implementations of existing systems (e.g., \cite{ZFQCL08}), but so too are better theoretical methods being proposed, such as the decoy state method \cite{Hwa03}.  Device-independent security proofs \cite{MY98,PABGMS09} aim to minimize the security assumptions on physical devices.  It seems reasonable to expect that further theoretical and engineering advances will eventually bring us devices which have strong arguments and few assumptions for their security.

\section{Key Usage: Encryption}\label{sec:enc}

The most commonly discussed usage for the key generated by quantum key distribution is encryption.  There are two ways \cite{PPS04} this key can be used for encryption.

In an \emph{unconditionally secure system}, the private key from QKD is used as the key in a one-time pad.  Since the key is information theoretically secure, so too is the encryption of the message: no computer, quantum or classical, will ever be able to decipher the encrypted message.  There are challenges to this system, however.  First, the one-time pad keys must be carefully stored and managed, as the double-use of one-time keys can seriously compromise security.  Second, as we discuss in Section~\ref{sec:limits}, physical QKD systems cannot yet achieve sufficiently high key generation rates to be able to encrypt large messages with one-time pads in real time.

To deal with this second challenge of low QKD key rates, \emph{hybrid systems} have been proposed, where the key from QKD is expanded with a classical stream cipher or block cipher such as the Advanced Encryption Standard (AES) to encrypt long messages.  In this setting, the security of the encrypted messages is no longer information theoretic: it depends on the computational assumption that the cipher used is hard to break.  While this is not ideal, it may not be too risky either.  Historically, cryptographers have been very good at designing block ciphers with few weaknesses: for example, the Data Encryption Standard (DES), designed in the 1970s, is no longer considered secure due to its short key length, but DES has stood up well to over 30 years of cryptanalytic attacks.  Under a known plaintext attack, the security of DES is reduced from $2^{56}$ to about $2^{41}$, but, when rekeying is sufficiently frequent, the effect of known plaintext attacks is limited \cite[\S 3.2]{SECOQC07}.  Moreover, quantum computers do not seem to have too much impact on ciphers: while Grover's search algorithm implies that the key length needs to be doubled, the exponentially faster attacks promised by Shor's algorithm and others do not apply to most ciphers.

Even when used in hybrid systems, QKD offers a substantial advantage over classical key agreement: the key from QKD is independent of any inputs to the key agreement protocol.  Thus, QKD reduces the number of points of attack: once a key has been established, the only way to attack such a system is to cryptanalyze the encryption.  By contrast, a system using classical key agreement could be attacked by trying to take the inputs to the classical key agreement protocol and determining the generated private key (e.g., by solving the Diffie-Hellman problem).  However, when using QKD to generate short keys, care must be taken due to finite length effects \cite{CS09}.

Hybrid QKD systems offer enhanced security compared to ciphers used without QKD: the QKD subsystem provides fresh, independent keying material frequently, which can rekey the classical block or stream cipher; with frequent rekeying, we reduce the risk of attacks against the underlying cipher that make use of many plaintexts or ciphertexts encrypted under the same key.

\section{Authentication}\label{sec:auth}

Quantum key distribution does not remove the need for authentication: indeed, authentication is \emph{essential} to the security of QKD, for otherwise it is easy to perform a man-in-the-middle attack.  There are two main ways to achieve authentication: public key authentication and symmetric key authentication.  \emph{Symmetric key authentication} can provide unconditionally secure authentication, but at the cost of needing to have pre-established pairs of symmetric keys.  \emph{Public key authentication}, on the other hand, is simpler to deploy, and provides extraordinarily convenient distributed trust when combined with certificate authorities (CAs) in a public key infrastructure (PKI).  Public key authentication cannot itself be achieved with information theoretic security.  We argue, however, that the security situation is more subtle than this: the use of public key authentication can still lead to systems that have very strong long-term security.

A third method for authentication is to use trusted third parties which actively mediate authentication between two unauthenticated parties, but there has been little interest in adopting these in practice.  Certificate authorities, which are used in public key authentication, are similar to trusted third party authentication but do not actively mediate the authentication: they distribute signed public keys in advance but then do not participate in the actual key authentication protocol.  The difference in trust between trusted third parties and certificate authorities for authentication in QKD is smaller than in the purely classical case since the key from QKD is independent of the inputs.

\subsection{Symmetric Key Authentication}\label{sec:auth:sym}

Parties who already share a short private key can use an unconditionally secure message authentication code to authenticate their messages.  The first such approach was described by Wegman and Carter \cite{WC81} and has been refined for use in QKD (for example, \cite{PNMLSPUFZ05}).  It is for this reason that quantum key distribution is sometimes called \emph{quantum key expansion}: it can take a short shared key and expand it to an information-theoretically secure long shared key.

Interestingly, the universal composability of quantum key distribution implies that we can use some of the key generated by QKD to authenticate the messages in the next round of QKD with a negligible decrease in security.  Thus we can continue QKD (more or less) indefinitely having started only with a relatively short (on the order of a few kilobytes) authentication key.

\subsection{Public Key Authentication}\label{sec:auth:pub}

While symmetric key authentication promises unconditionally secure authentication, it is difficult to deploy because each pair of communicating parties must share a private key.  Public key infrastructures allow for distributed trust and have been essential to the success of electronic commerce.  While many advocates of quantum cryptography dismiss the role of computationally secure public key authentication in QKD, we argue that public key authentication will be vital in a quantum key distribution infrastructure and can still provide meaningful security statements.

Public key authentication schemes, being computationally secure, tend to be broken, and invariably sooner than we expect.  In 1977, Rivest speculated \cite{Gar77} that it could take 40 quadrillion years to solve the RSA-129 problem (factoring a 129-decimal-digit RSA modulus), but it was broken only 17 years later \cite{AGLL94}.  While the popular press still occasionally uses expressions such as ``more than a quadrillion years'' \cite{Lys08} to describe the security of number-theoretic schemes, technical recommendations \cite{NISTSP80057,ECR08} are more nuanced and tend not to speculate too far beyond 2030.  Notably, these recommendations tend to ``assume [...] (large) quantum computers do not become a reality in the near future'' \cite[p. 25]{ECR08}.

Large scale quantum computers are widely believed to be some time off, but there appears to be no reason at present to doubt their eventual efficacy.  Quantum computers, however, are not the only threat against public key authentication.  Computers do become faster and new algorithms do help speed cryptanalysis.  However, we are not so pessimistic to think that all public key authentication is doomed forever.  In fact, we believe that public key authentication will continue to play a vital role in communication security indefinitely, even in the presence of quantum computing.

Although today's popular public key schemes --- RSA, finite field discrete logarithm, and elliptic curve --- would be broken by a large scale quantum computer, other ``post-quantum'' schemes do not immediately fall to quantum algorithms, and other schemes are sure to be developed ({\it cf.} \cite{BBD09}).  It seems to us, then, that public key schemes in the future are likely to go through a lifecycle in which a new primitive is proposed, it appears secure against current attack techniques, reasonable parameter sizes are proposed, adopted, and then computing technology and cryptanalysis advances chip away at the security until a newer scheme provides better tradeoffs.  It is not too hard to imagine a 20-year window in which a public key scheme, along with a particular set of parameter sizes, is considered viable.

It is in this scenario, where a particular public key authentication scheme is only deemed to be secure for a 20-year period, that quantum key distribution can thrive.  A public key authentication infrastructure provides the large scale usability that we have come to expect from PKIs, and when combined with quantum key distribution can offer strong security promises.  In quantum key distribution, the authentication --- in the form of public key authentication --- only needs to be secure up to and including the initial connection.  Once the QKD protocol has output some secret key, a portion of this secret can subsequently be used for symmetric key authentication.  In fact, even if the original authentication keys are revealed after the first QKD exchange, the key from QKD remains information theoretically secure.  In other words, we have the following statement:

\begin{quote}
If authentication is unbroken during the first round of QKD, \emph{even if it is only computationally secure}, then subsequent rounds of QKD will be information-theoretically secure.
\end{quote}

By contrast, classical public key exchange schemes do not have this feature.  Although one can employ a protocol in which a new key is transmitted encrypted under the old key, an eavesdropper who logs all communications and subsequently breaks the first key can read all future communications.  With QKD, new session keys are completely independent of all prior keys and messages.

\section{Limitations}\label{sec:limits}

Two undeniable limitations of present quantum key distribution schemes are distance and key rate.  Because of the fragile nature of the quantum mechanical state that is transmitted during quantum key distribution, the longer the distance that the photons have to travel, the more photons that are lost to decoherence and noise and hence the lower the rate of secret key formation.  Distance and key rate are a tradeoff, but progress is being made on improving the overall tradeoff.

\paragraph{Distance.}
The longest QKD experiments to date have acheived secure key generation over a 184.6km fiber optic link \cite{HRPHNLMN06} and over a free-space link spanning a distance of 144km at a rate of 12.8 bits/second\cite{SWFUTSPSK07} .  This free-space distance is considered sufficient to communicate between any two points on the surface of the Earth via orbiting satellites, the feasibility of which is the subject of a proposed experiment \cite{PFMMCPAJU08}.  

Quantum repeaters \cite{BDCZ98} would also overcome the distance limitation, allowing shared quantum states to be established between distant parties.  While these systems are not yet operational, they are easier to implement than full-scale quantum computers; theoretical and experimental work progresses on their development.

\paragraph{Key rate.}
While long distance experiments achieve very low key rates on the order of a few bits per second, shorter distance experiments have demonstrated very high key rates.  Experimental groups have achieved key rates of over 4 MB per second over 1km of fibre \cite{NISTQKD} and 1 Mb per second at 20km \cite{DYDSS08}.  These key rates are an impressive accomplishment are coming closer to the rates needed to secure real communication channels.

When a QKD key is used for encryption, current key rates may not be sufficient for a one-time pad and hybrid schemes need to be used, in which the QKD key is used as the private key in a symmetric encryption algorithm such as the block cipher AES.  However, as we have argued in Section~\ref{sec:enc}, even hybrid QKD systems offer enhanced security compared to classical key agreement since the keys generated by QKD are independent of any inputs to the key agreement procedure and since many symmetric encryption algorithms are resistant to known attacks by quantum computers.  Key rate can always be negatively impacted by an adversary disturbing the quantum channel, but such an adversary can not impact the security of the key agreement.

\section{QKD Networks}\label{sec:networks}

As QKD technology progresses, the structure of deployed QKD systems will progress in four stages to reduce distance limitations and increase commercial applicability:
\begin{enumerate}
\item \emph{Point-to-point links:} Two QKD devices are directly connected over a relatively short distance.
\item \emph{Networks with optical switches:} Multiple QKD devices are arranged in a network with optical switches to allow different pairs of interaction.  Optical switches, however, do not increase communication distance.  The switches need not be trusted.  One example of such a network is the DARPA quantum network \cite{ECPPSY05}.
\item \emph{Networks with trusted relays:} Multiple QKD devices are arranged in a network.  Intermediate nodes in the network can act as classical relays which relay information between distant nodes.  The relay nodes need to be trusted, although trust can be reduced by having the sender use a secret sharing scheme \cite{BS08}.  This type of QKD network would be suitable for scenarios where the operator of the network is also the user of the network, for example, a bank creating a network among its many branches, each of which is individually trusted.  One example of such a network is the SECOQC quantum network \cite{SECOQC07,SPDALL09}.
\item \emph{Fully quantum repeater network:} Multiple QKD devices are arranged in a network with quantum repeaters \cite{BDCZ98}.  Although individual links are still distance-limited, the quantum repeater nodes allow entanglement to be linked across longer distances, so QKD can be performed between distant parties.  The quantum repeaters need not be trusted, and this type of QKD network corresponds to the service provider scenario.
\end{enumerate}

\section{Conclusion}\label{sec:concl}

Quantum key distribution makes use of the eavesdropper-detection power offered by quantum mechanics to establish a shared key that is verifiably secure and independent of any other data, provided the communicating parties share an authentic channel.  The security of the system depends on no computational assumptions and thus has the potential to offer security against present or future attackers with unbounded classical or quantum computational power.

There are many scenarios, such as government, military, and health care, in which information needs to remain secure for 25, 50, or even 100 years.  Using QKD reduces the assumptions about the cryptographic system and produces a shared secret key that, by the properties of quantum mechanics, is independent of any other data, including the input.

It is important to consider how QKD fits into the larger cryptographic infrastructure.  When used with public key authentication, QKD provides strong security with the convenience of distributed authentication using public key infrastructures; the public key authentication scheme need only be secure up until QKD occurs, but the key from QKD will remain secure indefinitely.  If public key authentication is not possible, shared secret authentication can still be used to give enhanced security compared to classical key expansion.  

The present limitations of QKD --- distance and key rate --- will be further mitigated as experimental research in QKD continues, and quantum repeaters promise fully quantum long distance networks.

We believe that, as the technology continues to improve, QKD will be an increasingly valuable tool in the cryptographer's toolbox for building secure communication systems.

\subsection*{Acknowledgements}

The authors gratefully acknowledge helpful discussions with Romain All\'{e}aume, Daniel J. Bernstein, Hoi-Kwong Lo, Alfred Menezes, and Kenny Paterson.  Research performed while D.S. was at the University of Waterloo.  D.S. was supported in part by an NSERC Canada Graduate Scholarship.  M.M. is supported by a Canada Research Chair.  The authors acknowledge funding from the Ontario Centres of Excellence (OCE), Canada's NSERC, QuantumWorks, MITACS, CIFAR, Ontario-MRI, and Sun Microsystems Laboratories.

\small 

\bibliographystyle{halphads}
\bibliography{/Users/dstebila/Bibliography/Library}

\end{document}